\documentclass[aps,prl,twocolumn,showpacs,amsmath,amssymb,groupedaddress]{revtex4}

\usepackage{graphicx}
\usepackage{dcolumn}
\usepackage{bm}

\begin{document}


\title{Anomalies in the NMR of Silicon: Unexpected Spin Echoes in a Dilute Dipolar Solid}


\author{A.\,E. Dementyev, D. Li, K. MacLean, S.\,E. Barrett}
\email[e-mail: ]{sean.barrett@yale.edu}
\affiliation{Department of Physics, Yale University, New Haven, Connecticut 06511}
\homepage[web: ]{http://opnmr.physics.yale.edu}



\date{\today}

\begin{abstract} 
NMR spin echo measurements of $^{29}$Si in Silicon powders have uncovered
a variety of surprising phenomena that appear to be independent of doping.  These surprises include long tails and even-odd asymmetry in 
Carr-Purcell-Meiboom-Gill (CPMG) echo trains, and anomalous stimulated echoes with several peculiar
characteristics. Given the simplicity of this spin system, these results, which to date defy explanation, 
present a new and interesting puzzle in solid state NMR.
\end{abstract}


\maketitle


In order to implement quantum computation (QC) 
based upon spins in semiconductors \cite{kane,privman,loss,vrijen,ladd}, a
detailed understanding of spin dynamics in these materials is required.
To this end, we carried out a series of NMR measurements
that were motivated by a simple question: what is the $^{29}$Si decoherence time ($T_{2}$) in Silicon?
Earlier NMR studies in Silicon
addressed other questions \cite{shulman,lampel,holcomb}.

We find that it is possible to detect the $^{29}$Si (4.67$\%$ natural abundance (n.a.), spin-$\frac{1}{2}$) NMR signals out to much longer times than was previously thought
possible, and so far, we have been unable to explain these results in terms of well-known NMR 
theory \cite{slichter}. Surprises in such a simple spin system appear brand new to NMR, 
and understanding their origin is of fundamental importance.  
In this paper, we describe the phenomena and recount 
tests we have made to explore possible explanations.

Two standard experiments that measure $T_{2}$ are reported.  
First, using the Hahn echo sequence 
(HE: $90_{X}$-$\left(\frac{\mathrm{TE}}{2}\right)$-$180_{Y}$-$\left(\frac{\mathrm{TE}}{2}\right)$-ECHO \cite{hahn}),
the measured decay, with $T_{2_{HE}} \approx 5.6$ msec, is in quantitative agreement with that expected for the
static $^{29}$Si-$^{29}$Si dipolar interaction. 
This decay mechanism is commonly encountered in solids, 
and a number of ingenious pulse sequences have been invented to manipulate the interaction Hamiltonian,  
pushing echoes out to times well beyond $T_{2_{HE}}$ \cite{slichter,solidecho,spinlock,magic,WAHUHA,MREV,haeberlen,MehringText}.  
A common thread running through those sequences is the use of multiple $90^{\circ}$ pulses, and pulses applied
frequently compared to $T_{2_{HE}}$, which refocus the homonuclear dipolar coupling. 
The same cannot be said about the second sequence that we used to measure $T_{2}$,
the Carr-Purcell-Meiboom-Gill sequence (CPMG:  
$90_{X}$-\big\{$\left(\frac{\mathrm{TE}}{2}\right)$
-$180_{Y}$-$\left(\frac{\mathrm{TE}}{2}\right)$-ECHO\big\}$^{repeat\: n-times}$ \cite{cpmg}). 
Specifically, the CPMG sequence is not expected to excite echoes beyond $T_{2_{HE}}$,
since $180^{\circ}$ pulses should not affect the bilinear homonuclear interaction.
This statement is exact in two important limits:
 either for unlike spins or for magnetically-equivalent spins.

Therefore, we were surprised to find that 
{\em CPMG echoes are detectable long after $T_{2_{HE}}$, and the 
echo peaks appear nearly identical in Silicon samples with very different dopings.}
This CPMG ``tail'' appears to be even larger at
low temperatures.  In addition, as the interpulse spacing (TE) is increased, the CPMG echoes develop a pronounced
``even-odd asymmetry'' (e.g., long after spin echo $\#1$ (``SE1") is in the noise, spin echo $\#2$ (``SE2") is clearly observable). 
Lastly, we show how an ``anomalous stimulated echo" is observed in this system, with several peculiar
characteristics.  

\begin{figure}
\includegraphics[width=3.4 in]{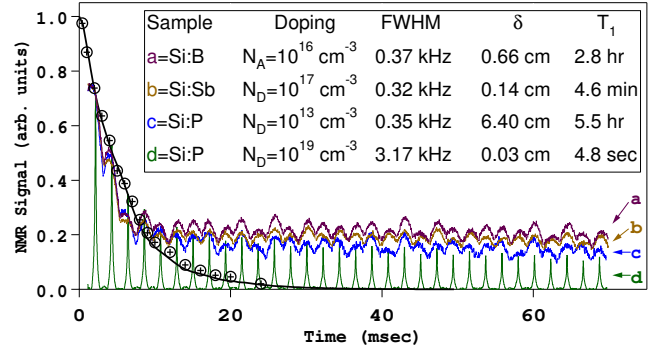}%
\caption{\label{fig1} Two standard measurements of the $^{29}$Si $T_{2}$ in powdered Silicon at room temperature (RT).
CPMG echo trains are shown for four samples (a-d) with different doping, full width half max (FWHM), skin depth 
($\delta$), and spin-lattice relaxation time ($T_{1}$). Since samples a, b, and c exhibit 
much wider echoes than sample d, only the top portion of their echoes are visible.
Hahn echo measurements (circles with crosses for sample d, others are supressed for clarity)
agree quantitatively with the dipolar decay curve (solid line) calculated for the Silicon lattice (Eqn. 4, see text).
Despite big changes in doping (e.g., $\times 10^{6}$ in P-concentration between samples c and d), 
the peaks of the CPMG echoes are nearly identical to each other, and they are detectable long after
the Hahn echoes decay to zero.
These measurements are in a 7.027 Tesla field 
$(\vec{B}\parallel\hat{z},$ with $ f_{o}=59.48$ MHz). N.B., all figures are in color online.}
\end{figure}

Figure 1 shows CPMG echo trains acquired in four different Silicon samples (both n-type and p-type).  As the legend
shows, the $^{29}$Si NMR spectrum (0.3 kHz$\le$FWHM$\le$3 kHz), the echo shape, and the
$T_{1}$ (from 4.8 sec to 5.5 hours at RT) can be quite different, for samples with wide variations 
in doping \cite{unpublished}. Despite these big changes (e.g., $\times 10^{6}$ in P-concentration), the peaks of the CPMG echoes are nearly identical to each other,
and they persist long after the Hahn echoes have died away.

Qualitatively, the long tail evokes a well-known effect in liquid-state NMR, where diffusion causes slow changes in
the local field leading to an extrinsic decay of the Hahn echoes \cite{slichter}. Applying frequent refocusing
pulses renders the dynamics ``quasi-static'', enabling the CPMG echoes to persist to longer times, and revealing the
intrinsic $T_{2}$.  However, in our data, the Hahn echoes appear to persist out to the ``intrinsic'' $T_{2}$ curve,
{\em and the CPMG echoes are observed beyond even that limit}, as we now show.

A theoretical decay curve may be calculated and compared to the experiments in Fig. 1, starting from a general 
spin Hamiltonian for $^{29}$Si in doped Silicon.  For example, for sample d, we have:
\begin{equation}
{\mathcal H}={\mathcal H}_{lab} + {\mathcal H}_{^{29}Si-^{29}Si} + {\mathcal H}_{^{29}Si-^{31}P} +  {\mathcal H}_{^{29}Si-e^{-}}, 
\label{eq1}
\end{equation} 
where ${\mathcal H}_{lab}$ includes the magnetic coupling of $^{29}$Si spins to both the
static laboratory field and the time-dependent tipping field produced by the rf pulses.  Since $^{29}$Si is fairly
dilute (4.67$\%$ n.a.), ${\mathcal H}_{^{29}Si-^{29}Si}$ is just the direct dipolar coupling.  
The last two terms, ${\mathcal H}_{^{29}Si-^{31}P}$ and ${\mathcal H}_{^{29}Si-e^{-}}$, play the role of the ``bath"
for the $^{29}$Si spins, which produce static magnetic shifts and determine $T_{1}$.  
In principle, the dynamics of this bath might also affect our $T_{2}$ measurements.   However, Fig. 1
shows that this is not the case, since samples a-d have nearly identical CPMG tails despite very different baths.
This is strong empirical evidence that the
$^{29}$Si homonuclear spin-spin coupling is sufficient to describe the physics of all four
samples (a-d), which greatly simplifies the model.
Therefore, in the rotating frame, the secular part of Eq. (\ref{eq1}) (in the absence of rf pulses)
is ${\mathcal H}_{r}$, given by \cite{slichter}:
\begin{equation}
\frac{{\mathcal H}_{r}}{\hbar}\!=\!\!\!\!\!\sum_{i}^{\!\!N_{spins}}\!\!\!\bigg(\!\!\Omega_{i}I_{z_{i}}\!\!\!+\!\!\!\!\!\!\sum_{j>i}^{N_{spins}}\!\!\!\!\Big\{\!a_{ij}I_{z_{i}}I_{z_{j}}\!\!\!+\!b_{ij}\!\left(I_{x_{i}}I_{x_{j}}\!\!\!+\!\!I_{y_{i}}I_{y_{j}}\!\right)\!\!\Big\}\!\!\bigg),
\label{eq2}
\end{equation} 
where $\Omega_{i}$ is the magnetic shift for spin i (relative to on-resonance spins),
$a_{ij}=\frac{(^{29}\gamma)^{2}\hbar}{r_{ij}^{3}} [1-3\cos^{2}\theta_{ij}]$ ($^{29}\gamma$ is the gyromagnetic
ratio for $^{29}$Si), and $b_{ij}=\frac{-a_{ij}}{2}$. The vector between spins i and j, ${\vec r_{ij}}$, satisfies
${\vec r_{ij}}\cdot{\hat z} =  r_{ij}\cos\theta_{ij}$.  

\begin{figure}
\includegraphics[width=3.4 in]{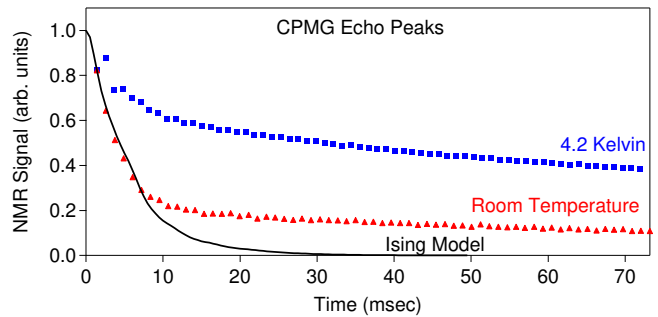}%
\caption{\label{fig2} CPMG echo peaks at RT (triangles) and 4.2 K (squares) for sample d. 
The sets are scaled so that SE1 agree.  The solid line is the calculated decay from Fig. 1 (see text).  
While the qualitative temperature effect is clear, the non-ideal conditions of the 
4.2 K data set (173$_Y$ pulses and repetition time=100 sec $\approx{T_{1}\over 3}$)
prevent a quantitative assessment.}
\end{figure}

If some of the terms in Eq. (\ref{eq2}) are truncated, corresponding to specific physical limits, then analytic
solutions  for the effect of various pulse sequences may be found using the product operator formalism
\cite{sorensen,slichter}. We start from the initial equilibrium density matrix:
\begin{equation}     
\rho(t=0) \propto\sum_{i}^{\!\!N_{spins}}I_{z_{i}},
\label{eq3}
\end{equation} which assumes the conventional strong field and high temperature  approximations \cite{slichter}.

For ``unlike spins'', where $|a_{ij}|\ll\Delta\Omega_{ij}\equiv|\Omega_{i}-\Omega_{j}|$, we truncate the $b_{ij}$
terms \cite{slichter}. In this limit, the peak of the $k$th CPMG echo decays according to: 
\begin{equation}     
\langle I_{Y}(k\!\times\!\mathrm{TE})
\rangle=\!\!\!\!\sum_{i}^{N_{spins}}\!\!\!I_{y_{i}}(0)\!\!\left\{\!\prod_{j>i}^{N_{spins}}\!\!\!\!\cos\!\left(\!{a_{ij}(k\!\times\!
\mathrm{TE})
\over 2}\!\right)\!\!\right\}\!, 
\label{eq4}
\end{equation} which assumes the ``infinite $H_1$ limit''.  Experimentally, $\frac{^{29}\gamma
H_1}{2\pi}$$\approx$$22 \mathrm{kHz}$, $|\frac{a_{ij}}{2\pi}|$$<$0.8 kHz, $|\frac {\Omega_{i}}{2\pi}|$$<$0.3 kHz
for samples (a-c) and $|\frac {\Omega_{i}}{2\pi}|$$<$3 kHz
for sample (d).
Eq. (\ref{eq4}) also describes a free induction decay (FID) following a single 90$_{X}$ pulse in another limit: all the $b_{ij}$ terms are
truncated {\em and} all $\Omega_{i}=0$.  Thus, the truncated dipolar decay of the CPMG echoes for the case of
``unlike spins'' is apparently unaffected by the 180$_{Y}$ pulses, which flip all $I_{z_{i}} \rightarrow
(-I_{z{i}})$, leaving the sign of the bilinear $a_{ij}$ terms unchanged.  In order to compare Eq. (\ref{eq4}) to
the data, we only need to have realistic values of $a_{ij}$ for our powder samples.  To obtain these 
$a_{ij}$, we
built 20,000 ``chunks" of the real Silicon lattice with arbitrary orientations, and determined the $\sim$80 
nearest neighbors occupied according to the 4.67$\%$ n.a..  Averaging Eq. (\ref{eq4}) over all ``chunks" \cite{unpublished}
yields the black curve shown in Fig. 1, which
agrees remarkably well with the Hahn echo data points, but which fails to describe the measured CPMG echoes.

When CPMG experiments are carried out in liquids, 
the well-known echo modulation due to J-coupling between unlike spins
can be effectively turned off \cite{FreemanText}, if pulses are applied so frequently that
$\frac{1}{\mathrm{TE}}$$\gg$J$_{ij}$ and $\frac{1}{\mathrm{TE}}$$\gg$$\Delta\Omega_{ij}$.
Similarly, in our solid-state measurements, applying a 
CPMG sequence with frequent pulses (i.e., small TE) might 
push the system artificially into the ``like spin'' regime, where
$|a_{ij}|$$\gg$$\Delta\Omega_{ij}$.   
In that limit, all the terms of Eq. (\ref{eq2}) should be retained.  This
precludes an analytic solution, but numerical calculations of $\langle I_{Y}(t)\rangle$ can be carried out for
small numbers of spins, including the required ensemble averaging
\cite{unpublished}.  These calculations show that the initial decay of the CPMG echoes in that limit should be
$\approx {2\over 3}$ {\em faster} than Eq. (\ref{eq4}), 
which agrees with the well-known second moment expressions
\cite{slichter, vanvleck}.  Our data require another explanation.

Empirically, the long tail induced by the CPMG sequence has several interesting characteristics \cite{unpublished}. 
For example, Fig. 2 shows that the tail height, relative to the 1st echo, grows as the sample is cooled down to 4.2 K.  This
result is only qualitative, since the tail height can also be changed by using a repetition time $<T_{1}$ and by
using tip angles slightly away from 180 degrees, which can be difficult to avoid at T=4.2 K. Still, even taking
these factors into account, the tail appears to be more pronounced at low temperatures \cite{unpublished}.

\begin{figure}
\includegraphics[width=3.4 in]{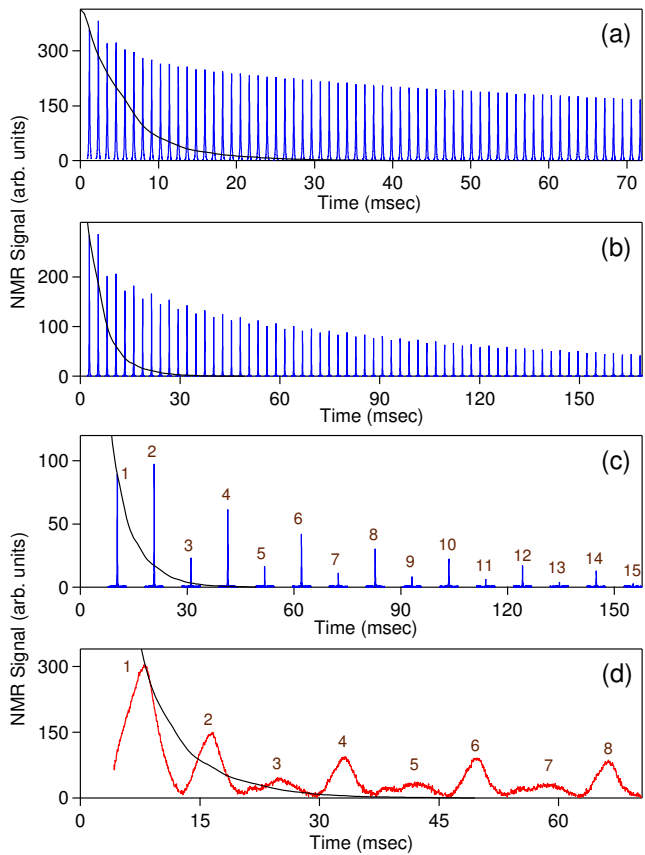}
\caption{\label{fig3} CPMG echo trains at 4.2 K for sample d with TE of (a) 1.12 msec, (b) 2.65 msec, and (c) 11.23
msec. The solid line from Fig. 1 is scaled to intercept the first echo in each graph.  
The numbered echoes in (c-d) exhibit a pronounced even-odd asymmetry, which emerges for TE $>T_{2_{HE}}$.
(d) shows the same effect in Si:Sb at RT with well-calibrated pulse angles,
low repetition rates, and a narrow spectrum.}
\end{figure}

To see if the long tail was due to some kind of multiple-pulse spin locking, we increased the interpulse
spacing (TE), which led to another unexpected result.  Figure 3(a-c) shows data
taken at 4.2 K for three different TE in sample d.   The long tail persists even for
TE$>T_{2_{HE}}$. 
Interestingly, for large interpulse spacings, the odd-numbered echoes are much smaller than the even-numbered
echoes.  At RT, samples a-d exhibit the same even-odd asymmetry as TE is increased (e.g., as in Fig. 3(d)).  

This even-odd asymmetry leads to remarkable results as TE is increased still further.  Figure 4 shows the FID 
and first two spin echoes acquired in a CPMG experiment with $n=2$, for very long TE.  
In Fig. 4(a) $\frac{\mathrm{TE}}{T_{2_{HE}}}\approx5.35$
so that SE1 is tiny relative to the FID.  Surprisingly, SE2 (at $\frac{2\times\mathrm{TE}}{T_{2_{HE}}}\approx10.7$) is
nearly three times the height of SE1; SE2 is also narrower than SE1. In Fig. 4(b), TE is doubled,  which pushes SE1
into the noise, while SE2 is clearly visible, {\em even though it occurs 21.4$\times$$T_{2_{HE}}$ after the 90$_{X}$
pulse}.  

\begin{figure}
\includegraphics[width=3.4 in]{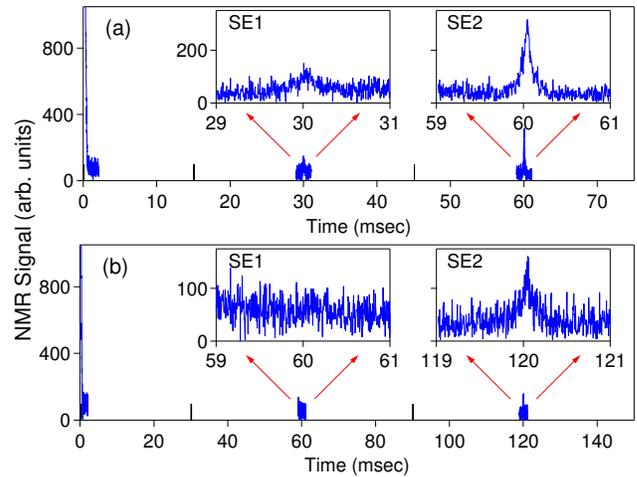}%
\caption{\label{fig4} 
The free induction decay (FID) and first two spin echoes (SE1, SE2) excited by a CPMG sequence at RT for sample d
 with TE of (a) 30 msec and (b) 60 msec.  The insets
show the narrow shape and the height of SE2 in comparison with SE1 
(the FID starts at 14600).  At 60 msec, SE2 (a) is clearly different from SE1 (b).  Solid bars indicate pulses.}
\end{figure}

Since SE2 is the first echo to occur after three pulses, we decided to look for a
contribution to the CPMG echoes that is reminiscent of a stimulated echo \cite{hahn}, using the sequence:
90$_{X}$-$\left(\frac{\mathrm{TE}}{2}\right)$-180$_{Y}$-TM-180$_{Y}$-DETECT, where TE and TM can be varied
independently.  Using this sequence, we detect a conventional spin echo SE2 that peaks at total time
$2\times\mathrm{TM}$, along with an ``anomalous stimulated echo'' (STE$_{A}$) that peaks at TM+TE.  Figure 5 shows the
height of the STE$_{A}$ as either TM or TE is varied.  There are several remarkable features of the data in Fig.
5:   1) we observe STE$_A$, even for our best 180$_{Y}$ pulses, where there should be none,
2) they decay slowly as TE or TM are increased, 3)
they appear to ``start''  at non-zero values at the left edge of Fig. 5, and 4) the data set has a larger scatter
than expected, given the signal to noise of each individual data point.

\begin{figure}
\includegraphics[width=3.4 in]{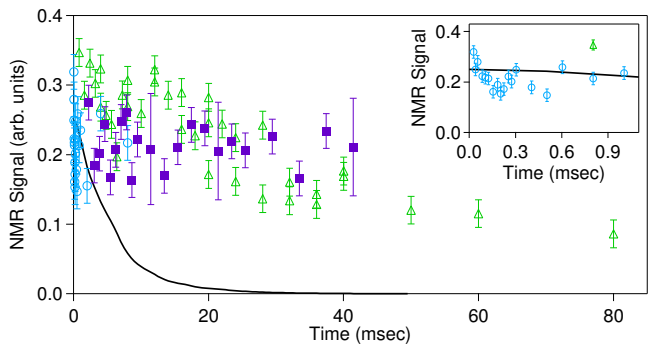}%
\caption{\label{fig5} ``Anomalous stimulated echo'' amplitudes at RT for sample d. Filled squares (TE$\approx$0.4 msec) are plotted
vs. TE+TM. Empty circles (TM$\approx$10 msec) and triangles (TM$\approx$21 msec) are plotted vs TE.  
The solid line is from Fig. 1.  (Inset) The signal does not appear to grow
from zero.}
\end{figure}

Given the results in Figs. 1-5, we have tried to minimize the effects of nonidealities
commonly reported in multiple pulse NMR \cite{slichter,MehringText,haeberlen}: 
(i) inhomogeneous $H_1$, (ii) finite-size $H_1$, (iii) a ``spin locking'' effect, 
and (iv) phase transients \cite{haeberlen}.  For (i), the results are unchanged if we use a tiny ($\sim 6$\% )
coil filling factor, or samples of very different skin depths.  For (ii), the same effects are
seen in all samples, even though $H_1$/FWHM changes by factor of 10.  For (iii), similar results are
obtained with an alternating phase Carr-Purcell sequence, where 180 degree pulse phases
alternate between``$-X$" and ``$X$", even though the average $H_1$ is quite different from that of CPMG.
Finally, we expect that (iv) becomes less important as the number of pulses is reduced and their spacing
is increased, so we don't see how this could explain the puzzling results of Figs. 4-5.  

Taken together, these results strongly suggest that the effects are due principally to the $^{29}$Si homonuclear
dipolar coupling.  In that case, why is it so hard to find a quantitative explanation for the data?  The form of Eq.
(\ref{eq2}) for a clean Silicon sample is one problem, since many spins may have $|a_{ij}|\sim\Delta\Omega_{ij}$,
which make simulations \cite{unpublished} particularly challenging \cite{allerhand}.  The dilution of the moments on
the lattice could be another issue \cite{lacelle}.  The strange features in Fig. 5 (in particular, point 3), and the
narrowness of SE2 in Fig. 4 seem to be beyond the conventional theory of solid-state NMR.  In recent NMR experiments
\cite{warren}, large polarizations have produced measurable dipolar field effects, which led some to question the
approximations underlying Eq. (\ref{eq3}).  While we don't have such large polarizations, the effects do appear
to be more pronounced at low temperatures (Fig. 2).

In the broader context of QC, the generic form of Eq. (\ref{eq2}) suggests that similar surprises may be found
in other systems with small, long-range, qubit-qubit interactions, particularly when ``bang-bang'' control is
used \cite{Chitty-Chitty}. Understanding these phenomena in Silicon may help
to prevent similar surprises from imposing a performance limit on quantum computers in the future.

This work was supported by the National Security Agency (NSA) and Advanced Research and Development Activity (ARDA)
under Army Research Office (ARO) contract $\#$DAAD19-01-1-0507. We would like to thank T.P. Ma (Yale EE) for
providing the Si:P samples,  and also R. Falster (MEMC) for providing the  Si:B and Si:Sb samples that were used in
this study.  We thank C.P. Slichter, R. Tycko, W.S. Warren, D.G. Cory, T.F. Havel, J.S. Waugh, R.G. Griffin, S.M.
Girvin, A. Vestergren, and V.V. Dobrovitski for stimulating discussions.  In particular, we appreciate the help of K.W. Zilm and X.
Wu, who independently verified several of these results.

\end{document}